\documentclass[aps,
preprint,
superscriptaddress]{revtex4}

\usepackage{amsfonts}
\usepackage{amsmath}
\usepackage{amssymb}
\usepackage{graphicx}%
\usepackage{hyperref}
\usepackage[dvipsnames]{xcolor}
\usepackage{ulem}

\begin{document}



\title{Transition from Pauli paramagnetism to Curie-Weiss behavior in vanadium}



\author{A. S. Belozerov}
\affiliation{M. N. Mikheev Institute of Metal Physics, Russian Academy of Sciences, 620108 Yekaterinburg, Russia}

\author{A. A. Katanin}
\affiliation{Center for Photonics and 2D Materials, Moscow Institute of Physics and Technology, 141701 Dolgoprudny, Russia}
\affiliation{M. N. Mikheev Institute of Metal Physics, Russian Academy of Sciences, 620108 Yekaterinburg, Russia}

\author{V. I. Anisimov}
\affiliation{M. N. Mikheev Institute of Metal Physics, Russian Academy of Sciences, 620108 Yekaterinburg, Russia}
\affiliation{Institute of Physics and Technology, Ural Federal University, 620002 Yekaterinburg, Russia}

\begin{abstract}
We study electron correlations and their impact on magnetic properties of bcc vanadium by a combination of density functional and dynamical mean-field theory. The calculated uniform magnetic susceptibility {in bcc structure} is of Pauli type at low temperatures, while it obeys the Curie-Weiss law at higher temperatures. Thus, we qualitatively reproduce the experimental temperature dependence of magnetic susceptibility without introducing the martensitic phase transition. Our results for local spin-spin correlation function and local susceptibility reveal that the Curie-Weiss behavior appears due to partial formation of local magnetic moments, which originate from $t_{2g}$ states and occur due to local spin correlations caused by Hund's rule coupling. At the same time, the fermionic quasiparticles remain well-defined, while the formation of local moments is accompanied by a deviation from the Fermi-liquid behavior. In particular, the self-energy of the $t_{2g}$ states shows the non-analytic frequency dependence, which is a characteristic of the spin-freezing behavior, while the quasiparticle damping changes approximately linearly with temperature in the intermediate temperature range $200$--$700$~K. By analyzing the momentum dependence of static magnetic susceptibility, we find incommensurate magnetic correlations, which may provide a mechanism for unconventional superconductivity at low temperatures.
\end{abstract}

\maketitle

\section{Introduction}

The $d$-electrons of transition metals may show both itinerant and localized behavior, that complicates the construction of unified theory.
In particular, the itinerant magnetism of transition metals can be described by the spin-fluctuation theory of Moriya~\cite{Moriya_book}, which almost reproduces the Curie-Weiss law for weak ferromagnetic materials. 
At the same time, a description of systems with well-defined local magnetic moments is usually performed by Heisenberg-type models.

Formation of local magnetic moments in the metallic state can be induced by electronic correlations, which may yield the so-called ``spin freezing'' (i.e., local moment) behavior found
in the three-band Hubbard model~\cite{spin_freezing} and in a number of real materials~\cite{OurAlpha0,OurAlpha,Belozerov2013,Pourovskii,Sangiovanni,OurGamma,Toschi,OurPnic}.
The spin-freezing phase is characterized by fulfillment of the Curie-Weiss law for the static local susceptibility, that indicates the presence of local magnetic moments.
The transition to
this phase was shown to be accompanied by the non-Fermi-liquid behavior of electronic self-energy~\cite{spin_freezing, our_dos_asymmetry}.
While the case of ``strong" magnets with almost formed local moments was intensively studied \cite{Licht_Kats_Grechnev_Benea,OurAlpha0,OurAlpha,Belozerov2013,Pourovskii,Sangiovanni,leonov_fe}, only some results where obtained for materials with partly (and weakly) formed local magnetic moments \cite{OurPnic,OurGamma,epsilon_fe,OurZrZn2}. These latter substances represent an interesting limit of almost itinerant systems, 
in which local moments appear at elevated temperature, but they are destroyed at low temperatures, where quasiparticle coherence occurs.
%

In this paper, we consider
metallic vanadium, a paramagnet with body-centered cubic (bcc) lattice, whose magnetic susceptibility is almost temperature-independent below 250 K, while above this temperature it obeys the Curie-Weiss law~\cite{susc1_burger61,susc2_suzuki65}.
Performing thermal-expansion measurements, Bollinger \textit{et al} reported a structural distortion below temperature of 300~K~\cite{Bollinger2011}, which is inconsistent with cubic crystal structures~\cite{Jung1977,Westlake1967}.
%
In principle, such a transition could explain the crossover from Pauli to the Curie-Weiss behavior in vanadium.
However, a presence of this martensitic phase transition in V was later questioned~\cite{Cordero2018} and obviously an additional verification by more sensitive experimental techniques is required. The natural question is also whether such a transition is necessary to explain the crossover from Pauli to Curie-Weiss behavior.

Other properties of vanadium are also intriguing.
In particular, 
it was revealed to be 
a type~II superconductor with the critical temperature of 5~K~\cite{Wexler1952}.
At the temperatures 
between 160 
and 270~K the
anomalies 
of
temperature dependencies of electrical resistivity  ~\cite{Rostoker1955,White1959,Hren1960}, elastic constants~\cite{Bolef1971_1972},  lattice parameters~\cite{Smirnov1966,suzuki66}, and thermoelectric power~\cite{Mackintosh1963} have been reported.
%
These anomalies were proposed to arise due to onset of antiferromagnetic ordering~\cite{susc1_burger61,Smirnov1966}, which could also contribute to the superconducting pairing, as it was emphasized in Ref.~\cite{Mackintosh1963}.
However, neutron diffraction~\cite{Shull1953} and nuclear magnetic resonance~\cite{Drain1964} studies did not find any static magnetic order.
As suggested by Westlake~\cite{Westlake1967}, some of the observed anomalies may be attributed to hydrogen content in samples.
Indeed, a subsequent study of dehydrogenated samples confirmed an absence of anomalies in elastic constants and electrical resistivity~\cite{Westlake1972}. Yet, presence of antiferromagnetic correlations in vanadium remain an open question. Also, the resistivity depends linearly on temperature in the intermediate temperature range 100--300~K \cite{susc1_burger61,Westlake1967,Westlake1968}. Although this temperature dependence was explained by contribution of phonons \cite{Westlake1968}, an important question is whether local and magnetic correlations contribute to this behavior.

%
%

%

Theoretical studies of vanadium were performed
by band structure methods with electronic potentials adjusted to experimental data~\cite{Wakoh1973_1975},
local density approximation (LDA)~\cite{Nirmala2003} and LDA+$U$ approach~\cite{Tokii2003}.
These calculations reproduced the experimental pressure dependence of superconducting transition temperature~\cite{Nirmala2003}, as well as shapes and dimensions of Fermi surface hole pockets around N-point~\cite{Tokii2003} found by using de~Haas-van Alphen effect~\cite{dHvA}.
At the same time, the momentum distributions obtained by Compton scattering were well described only at moderate and large momentum~\cite{Wakoh1973_1975}.

Transition metals often show substantial many-body effects
due to partially filled electronic subshells.
These effects require a careful treatment within density functional theory (DFT), which can be achieved, e.g., by using a sophisticated exchange-correlation functional,
avoiding symmetry restrictions~\cite{DFT_symmetry_restrictions} or combining with a disordered local moment (DLM) model~\cite{dlm_method}.

Another approach is to combine DFT with the Hubbard model, 
which can be solved via the static mean-field approximation (LDA+$U$ method~\cite{Anisimov1991}) or dynamical mean-field theory (DMFT)~\cite{dmft}.
The latter combination, called DFT+DMFT~\cite{dftdmft}, 
allows one to take into account local spin and charge dynamics, as well as temperature effects.
Previous studies of vanadium by DFT+DMFT method
revealed significant electron correlations
and provided a remarkably better
agreement with experimental Fermi surface~\cite{dmft_weber17} and spectroscopic data~\cite{Sihi2020}
than LDA and LDA+$U$ approaches.
%

In this paper, we perform a DFT+DMFT study of the electronic and magnetic properties of bcc vanadium.
The transition from Pauli paramagnetism to Curie-Weiss law is qualitatively reproduced in bcc structure.
We also demonstrate that electron correlations and, in particular, the Hund's exchange interaction are responsible for \textit{partial} formation of local magnetic moments at high temperatures.

\vspace{0.4cm}
\section{Method and computational details} \label{sec:computational_details}

We employ a fully charge self-consistent DFT+DMFT approach~\cite{charge_sc} implemented with plane-wave pseudopotentials~\cite{espresso,Leonov1}.
The Perdew-Burke-Ernzerhof form of generalized gradient approximation was used.
For a bcc lattice of vanadium we adopt the experimental lattice constant of 3.028~\AA~\cite{dmft_weber17}.
However, we checked that calculations with equilibrium lattice constant of 3.061~\AA\ obtained in DFT+DMFT give rise to qualitatively similar results.
The convergence threshold for total energy was set to $10^{-6}$~Ry.
The kinetic energy cutoff for wavefunctions was set to 60~Ry.
The reciprocal space integration was performed using ${30\times 30\times 30}$\, ${\bf k}$-point grid except the calculations of momentum-dependent susceptibility, where ${60\times 60\times 60}$ grid was used.

To take into account the on-site Coulomb correlations, we consider the following Hamiltonian within DMFT approach~\cite{dmft}:
\begin{equation}
\hat{H}_{\rm DMFT} = \hat{H}_{\rm DFT}^{\rm WF} + \hat{H}_{\rm Coul} - \hat{H}_{\rm DC},
\end{equation}
where $\hat{H}_{\rm DFT}^{\rm WF}$ is
the effective low-energy Hamiltonian constructed as a projection of Kohn-Sham states obtained within DFT onto the basis of Wannier functions, 
$\hat{H}_{\rm Coul}$ is the local Coulomb interaction Hamiltonian,
and $\hat{H}_{\rm DC}$ is the double-counting correction for accounting of the on-site interaction already described by DFT.
In our study, we compute $\hat{H}_{\rm DFT}^{\rm WF}$ self-consistently with respect to the redistribution of charge density caused by electronic correlations in DMFT part, as described in Ref.~\cite{charge_sc}.
For construction of $\hat{H}_{\rm DFT}^{\rm WF}$, we use a basis set of atomic-centered Wannier functions within the energy window spanned by the $3d$-$4s$ band complex~\cite{Wannier}.

We consider the local Coulomb interaction in the density-density form,
\begin{equation}
\hat{H}_{\rm Coul} = \frac{1}{2}\sum\limits_{mm'\sigma\sigma'} U_{mm^\prime}^{\sigma\sigma^\prime}
\hat{n}_{m\sigma} \hat{n}_{m^\prime\sigma^\prime},
\end{equation}
where $\hat{n}_{m\sigma}$ is the particle number operator at orbital $m$ with spin $\sigma$,
$U_{mm^\prime}^{\sigma\sigma^\prime}$ is the local interaction matrix.
For a parametrization of $U_{mm^\prime}^{\sigma\sigma^\prime}$, we use Slater integrals $F^0$, $F^2$, and $F^4$ linked to the Hubbard parameter ${U\equiv F^0}$ and Hund's rule coupling ${J_{\rm H}\equiv (F^2+F^4)/14}$ 
(see Ref.~\cite{u_and_j}).
We perform our calculations with ${U=2.3}$~eV and ${J_{\rm H}=0.9}$~eV, which were used in a previous DFT+DMFT study and led to a good agreement of the calculated Fermi surface with experimental data~\cite{dmft_weber17}.
However, available estimates of $U$ by the constrained random-phase approximation (cRPA) lie between 2.3 and 3.4~eV in our notations and depend on the choice of energy windows and bands~\cite{Aryasetiawan2006,Miyake_Sasioglu,Sihi2020}. Moreover, constrained density functional theory (cDFT) calculations resulted in even larger $U$ of about 4--5~eV~\cite{Aryasetiawan2006}.
Therefore, we also consider ${U=4}$~eV and ${U=6}$~eV for a more comprehensive study.

We use a double-counting correction in the around mean-field form~\cite{AMF}, $\hat{H}_{\rm DC} = [U \hat{n}_d (2 n_{\rm orb} {-} 1) - J_{\rm H} \hat{n}_d (n_{\rm orb} {-} 1)]/(2n_{\rm orb})$, where
$\hat{n}_d$ is the number of $d$ electrons, $n_{\rm orb}$ is the number of correlated orbitals.
%
%
We also verified that the fully localized form of double-counting correction leads to similar results with a slightly smaller (by $\sim$0.03) filling of \textit{d} states.

In DMFT, the lattice problem is mapped onto an effective impurity problem (Anderson model), in which the quantum impurity is embedded in an self-consistency determined electronic bath~\cite{dmft}.
To solve this quantum impurity problem within DMFT, we employ the hybridization expansion continuous-time quantum Monte Carlo method~\cite{CT-QMC},
which is used to simulate the following  impurity action:
\begin{equation}
S_{\rm imp} =- \int_0^\beta \! d\tau \int_0^\beta \! d\tau^\prime \sum_{m\sigma} c_{m\sigma}^\dag(\tau)\mathcal{G}_{0,m}(\tau-\tau^\prime)^{-1}\, c_{m,\sigma}(\tau^\prime) +  \frac{1}{2}
\int_0^{\beta}\! d\tau \sum\limits_{mm'\sigma\sigma'} 
U_{mm^\prime}^{\sigma\sigma^\prime}
n_{m\sigma}(\tau) n_{m^\prime\sigma^\prime}(\tau),
\end{equation}
where $\tau$ and $\tau^\prime$ are imaginary times, 
$c_{m\sigma}^\dag$ ($c_{m\sigma}$)
is the Grassmann variable for creation (annihilation) of an electron with spin~$\sigma$ at orbital~$m$,
$\mathcal{G}_{0,m}(\tau-\tau^\prime)$ is the bath Green function, which can be expressed via the Dyson equation. 
%
%
%
To perform the analytical continuation of obtained self-energies to real-energy range, we employ Pad\'e approximants~\cite{Pade}.

\section{Results and discussion}
\subsection{Electronic properties\label{Sect:elProp}}

Our DFT+DMFT calculations result in the $d$-states filling in the range 3.93--3.95 for all considered values of Hubbard $U$. The obtained filling is not far from half-filling and thus favorable for significant many-body effects~\cite{spin_freezing}. We also find that orbitals of $t_{2g}$ and $e_g$ symmetry have close fillings of about 0.83 and 0.71, respectively.

\begin{figure}[t]
\centering
\includegraphics[clip=true,width=0.52\textwidth]{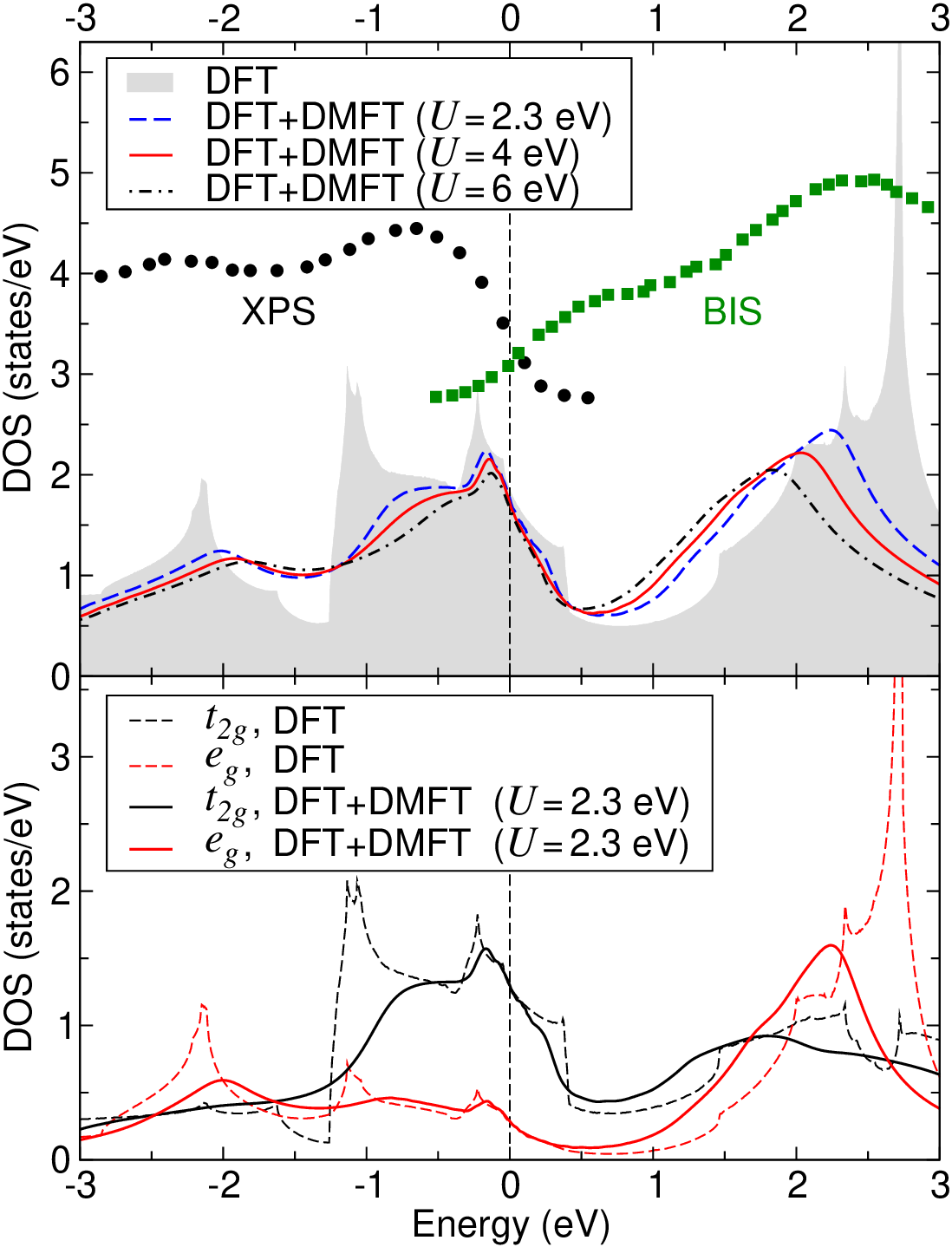}
\caption{
\label{fig:dos}
Total (top panel) and orbital projected (bottom panel) density of states obtained by non-magnetic DFT and DFT+DMFT at temperature ${T = 387}$~K with various values of Hubbard parameter $U$ and Hund's coupling ${J_{\rm H}=0.9}$~eV
in comparison with results of x-ray photoelectron spectroscopy
(XPS)~\cite{XPS} and bremsstrahlung isochromat spectroscopy (BIS)~\cite{BIS} shown in arbitrary units. The Fermi level is at zero energy.}
\end{figure}

In the top panel of Fig.~\ref{fig:dos}, we present the total density of states (DOS) computed by non-magnetic DFT and DFT+DMFT at temperature ${T = 387}$~K in comparison with experimental spectra~\cite{XPS,BIS}.
%
%
One can see a broadening of sharp peaks of DOS obtained within DFT due to account of temperature effects and local correlations in DMFT.
The latter also result in shifting of DOS features closer to the Fermi level, while keeping the value of DOS at the Fermi level.
The best agreement with presented experimental spectra is found at ${U=2.3}$~eV. Nevertheless, a slightly larger ${U\sim 3}$~eV is also expected to provide a reasonable agreement with experiment.
Our results also agree with spectral functions obtained in previous DFT+DMFT study~\cite{Sihi2020}.

All obtained DOSes have a peak just below the Fermi level, which may substantially enhance the electron correlations and thereby affect the magnetic properties, as found earlier in $\alpha$ iron~\cite{OurAlpha0} and model studies~\cite{our_dos_asymmetry}.
The distance from the peak to the Fermi level
is found to decrease monotonically from 0.16 to 0.12~eV with increase of $U$ from 2.3 to 6~eV.
To elucidate the origin of this peak, in the bottom panel of Fig.~\ref{fig:dos} we show the orbital projected DOS.
Although both $t_{2g}$ and $e_g$ states contribute to this peak, the former provide a dominant contribution.  
%
%

\begin{figure}[t]
\centering
\includegraphics[clip=true,width=0.53\textwidth]{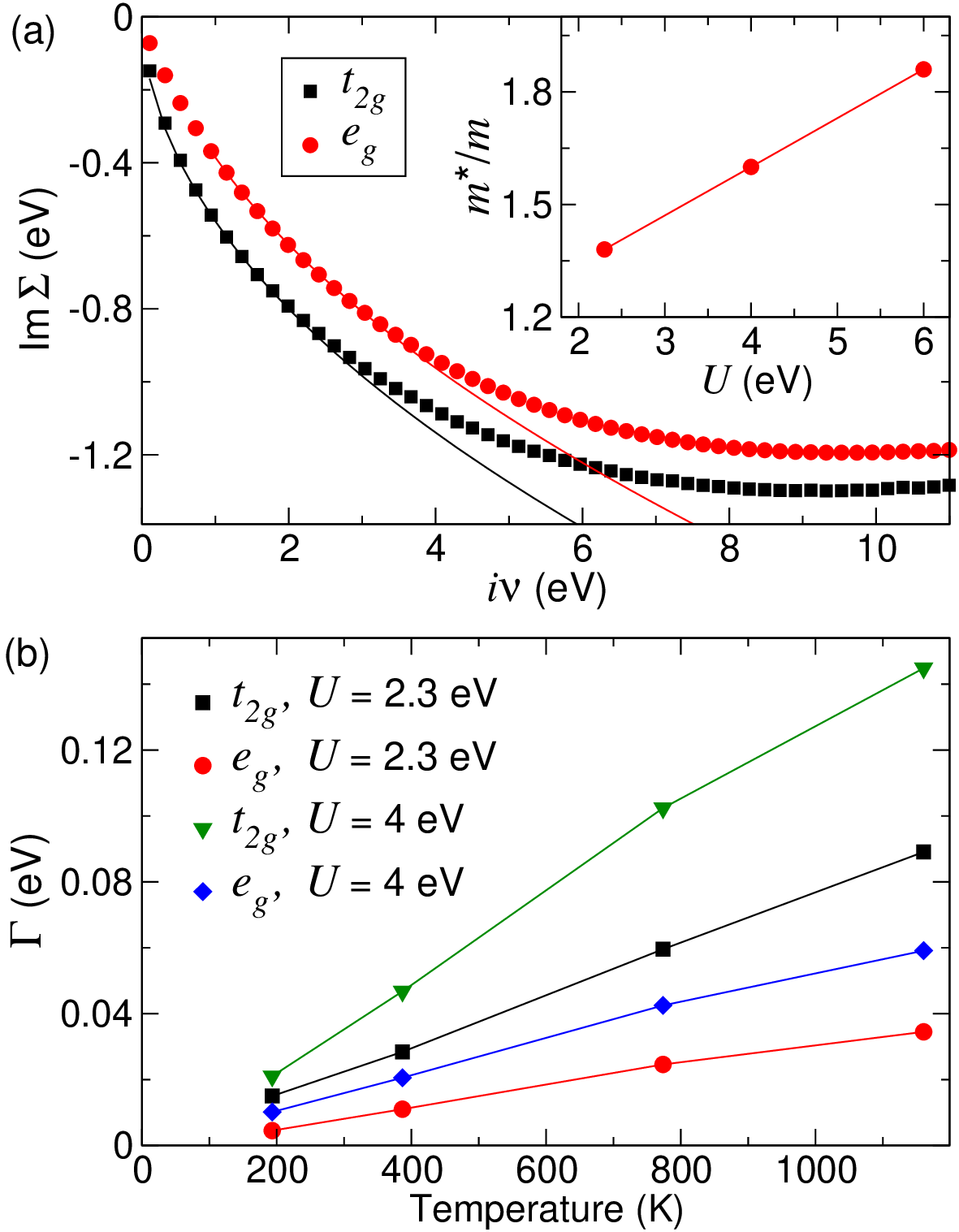}
\caption{
\label{Fig:sigma}
(a) Imaginary part of electronic self-energy (main panel) as a function of imaginary Matsubara frequency $i\nu$ obtained by DFT+DMFT method at temperature ${T = 387}$~K with Hubbard parameter ${U=6}$~eV and Hund's coupling ${J_{\rm H} = 0.9}$~eV. Thin lines show the fits to $A+B\nu^{1/2}$ dependence in the low (intermediate) frequency range for $t_{2g}$ ($e_g$) states. Inset: mass enhancement factor ${m^*/m}$ as a function of~$U$ at ${J_{\rm H} = 0.9}$~eV.
(b) Temperature dependence of quasiparticle damping $\Gamma$ with ${J_{\rm H} = 0.9}$~eV and two values of $U$. The lines are guides to the eye.
}
\end{figure}

For well defined electronic quasiparticles
the imaginary part of electronic self-energy
depends on small Matsubara frequency $\nu_n$ as ${\textrm{Im}\, \Sigma(i\nu_n)\approx -\Gamma -(Z^{-1} {-}1)\nu_n}$, where
$\Gamma$ is the 
damping, corresponding to the inverse quasiparticle lifetime, and
$Z$ is the 
quasiparticle residue. For $e_g$ states the obtained frequency dependence of self-energy has a 
quasiparticle shape at low frequencies at all considered values of $U$
(see, in particular, Fig.~\ref{Fig:sigma}a for ${U=6}$~eV). To estimate the strength of correlation effects, we also compute $Z^{-1}$ for $e_g$ states,
which is equal to the quasiparticle mass enhancement factor ${m^*/m}$ due to locality of self-energy in DMFT. 
%
%
%
The obtained values of ${m^*/m}$ (see inset of Fig.~\ref{Fig:sigma}a) are larger than those of 1.17 in chromium~\cite{ourChromium} and 1.25 in nickel~\cite{Sangiovanni}, 
and characterize $e_g$ states of vanadium as being moderately correlated. At the same time, the self-energy of $t_{2g}$ states is better described by $\nu^{1/2}$ frequency dependence, which is characteristic for the crossover to the spin-freezing state \cite{spin_freezing}. Similar dependence is also observed at intermediate frequencies for $e_g$ states.  
%
%
In the bottom panel of Fig.~\ref{Fig:sigma},
we present the obtained quasiprticle damping $\Gamma=|{\rm Im}\Sigma(\nu\rightarrow 0)|$ as a function of temperature. For the Fermi liquid $\Gamma$ is expected to depend quadratically on temperature. Instead, it shows almost linear dependence above ${T=200}$~K.
Therefore, the correlated state of vanadium in this temperature range cannot be completely described by the Fermi liquid concept. The obtained linear temperature dependence of the scattering rate may also contribute to the linear temperature dependence of resistivity (cf. also the recent single-band study of Ref. \cite{KataninMazitov}), such that its earlier proposed phonon-induced origin \cite{Westlake1968} should be reconsidered.

\subsection{Magnetic properties\label{Sect:magnProp}}

\begin{figure}[t]
\centering
\includegraphics[clip=true,width=0.61\textwidth]{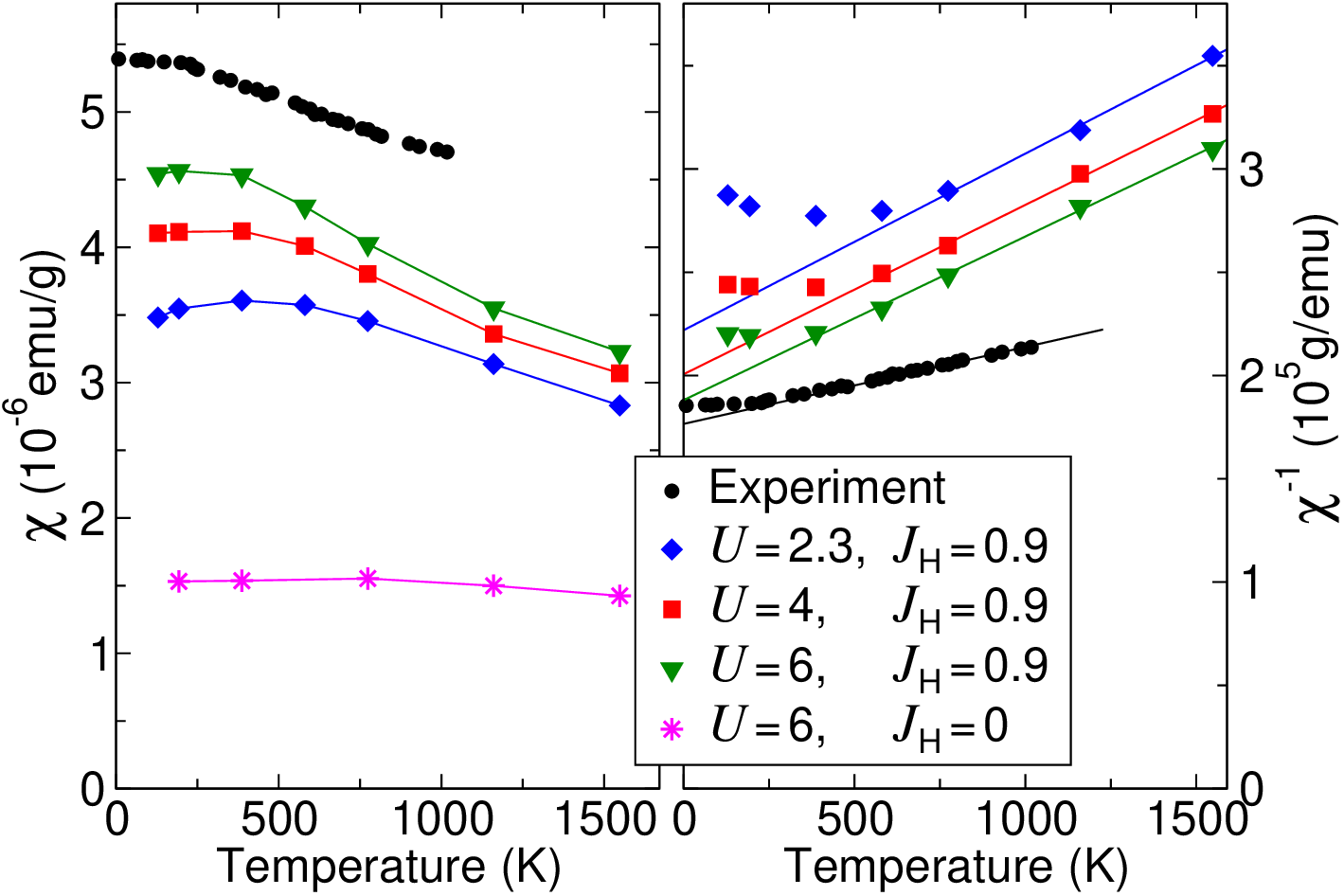}
\caption{
\label{Fig:chi_uniform}
Temperature dependence of the uniform magnetic susceptibility $\chi$ (left panel) and its inverse (right panel) calculated by DFT+DMFT method with various values of Hubbard $U$ and Hund's coupling $J_{\rm H}$ (in eV) in comparison with experimental data~\cite{susc2_suzuki65}. The straight lines at the right panel depict the least-squares fit to the linear dependence at temperature ${T>700}$~K.}
\end{figure}

First, we compute the uniform magnetic susceptibility as a response to a small external magnetic field. 
In particular, we used the magnetic field resulting in splitting of single-electron energies by 40~meV.
This field was checked to provide a linear response. 
In Fig.~\ref{Fig:chi_uniform} we present the calculated uniform magnetic susceptibility and its inverse in comparison with experimental data~\cite{susc2_suzuki65}.
The susceptibilities obtained with ${J_{\rm H}=0.9}$~eV qualitatively reproduce the Pauli-type behavior at low temperatures and obey the Curie-Weiss law at high temperatures.
%
%
Considering $U$ values from 2.3 to 6~eV, we find that an increase of $U$ increases the susceptibility magnitude and decreases the 
temperature of the onset of Curie-Weiss behavior $T_{\rm onset}$ from ${\sim700}$ to ${\sim300}$~K,
leading to a better agreement with experiment.
Fitting the inverse susceptibility at temperature ${T>700}$~K to the Curie-Weiss law
${\chi^{-1} =3(T{-}\Theta)/\mu^2_{\rm eff}}$,
we extract the effective local magnetic moment ${\mu_{\rm eff}=2.2~\mu_{\rm B}}$,
which is less than the experimental value of 3.3~$\mu_{\rm B}$ and is found to be weakly dependent on $U$.

To determine the role of Hund's coupling, we compute the uniform magnetic susceptibility with
${J_{\rm H} = 0}$.
As shown in Fig.~\ref{Fig:chi_uniform}, these calculations led to the suppression of the Curie-Weiss behavior even at relatively large ${U=6}$~eV.
Hence, the Curie-Weiss behavior in vanadium
is due to local spin correlations, controlled by Hund's exchange, rather than charge ones provided by~$U$.
Since the obtained results are more sensitive to ${J_{\rm H}}$ than to $U$, we additionally checked (not shown in figure) that the agreement with experiment may be further improved by using ${J_{\rm H}=1}$~eV, which is also commonly used for $3d$ subshell.

For the lowest considered value of $U$ in the presence of Hund exchange the temperature dependence of the uniform susceptibility shows a maximum and weak decrease with decreasing temperature in the low-temperature regime. While the maximum of the susceptibility reflects a crossover from itinerant (screened local moment, see below) to the unscreened local moment behavior, the increase of the susceptibility at low temperatures reflects increase of density states of thermally excited quasiparticles, which is provided by the offset of the peak of density of states from the Fermi level. Note that similar increase of the susceptibility with temperature in the low-temperature regime was obtained previously in $\gamma$-iron \cite{OurGamma} and iron pnictides \cite{OurPnic}.

\begin{figure}[t]
\centering
\includegraphics[clip=true,width=0.53\textwidth]{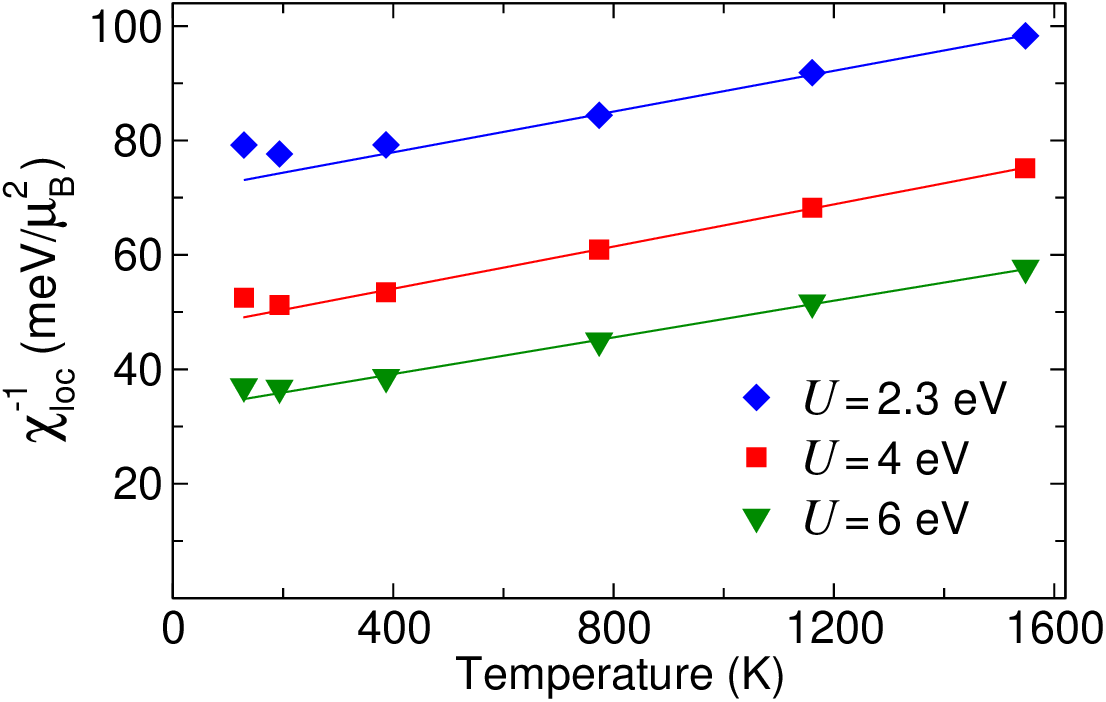}
\caption{
\label{Fig:chi_loc}
Temperature dependence of the inverse local magnetic susceptibility obtained by DFT+DMFT method
with ${J_{\rm H}=0.9}$~eV and different values of Hubbard $U$. The straight lines depict the least-squares fit to the linear dependence above 700~K.}
\end{figure}

To ensure that the Curie-Weiss law is caused by local spin correlations,
we calculate the local magnetic susceptibility $\chi_{\rm loc}=4\mu_{\rm B}^2 \int_0^\beta \langle S_z(\tau) S_z(0) \rangle d\tau$,
where $S_z$ is the $z$-component of the
local spin operator,
$\beta$ is the inverse temperature,
$\tau$ is the imaginary time.
The inverse of $\chi_{\rm loc}$, shown in Fig.~\ref{Fig:chi_loc}, also depends linearly on temperature above 400~K, which is of the order of $T_{\rm onset}$, implying formation of
local magnetic moments. We therefore connect the obtained Curie-Weiss law of the uniform susceptibility with the presence of partly formed local magnetic moments and associate the temperature $T_{\rm onset}$ with the Kondo temperature of screening of local magnetic moments ${T_{\rm K}\sim T_{\rm onset}}$. Note that while for well defined local magnetic moments ${T_{\rm K}\sim - T_{\rm W}}$, where the Weiss temperature $T_{\rm W}$ is defined by the Curie-Weiss law for the local susceptibility ${\chi^{-1}_{\rm loc} =(T{-}T_{\rm W})/C}$ \cite{Wilson,Melnikov,Tsvelik,Comment,Reply}, in the considered case, similarly to the previous study of weak itinerant magnet ZrZn$_2$ \cite{OurZrZn2}, we find $T_K\ll |T_W|$. 
As discussed in Sec.~\ref{Sect:elProp}, the Fermi liquid state is formed at 
the temperature ${T^*\lesssim 200}$~K$<T_K$.
Hence, the transition from Pauli to Curie-Weiss behavior may be associated with increasing loss of quasiparticle coherence and the respective local moment formation upon heating.

\begin{figure}[t]
\centering
\includegraphics[clip=true,width=0.61\textwidth]{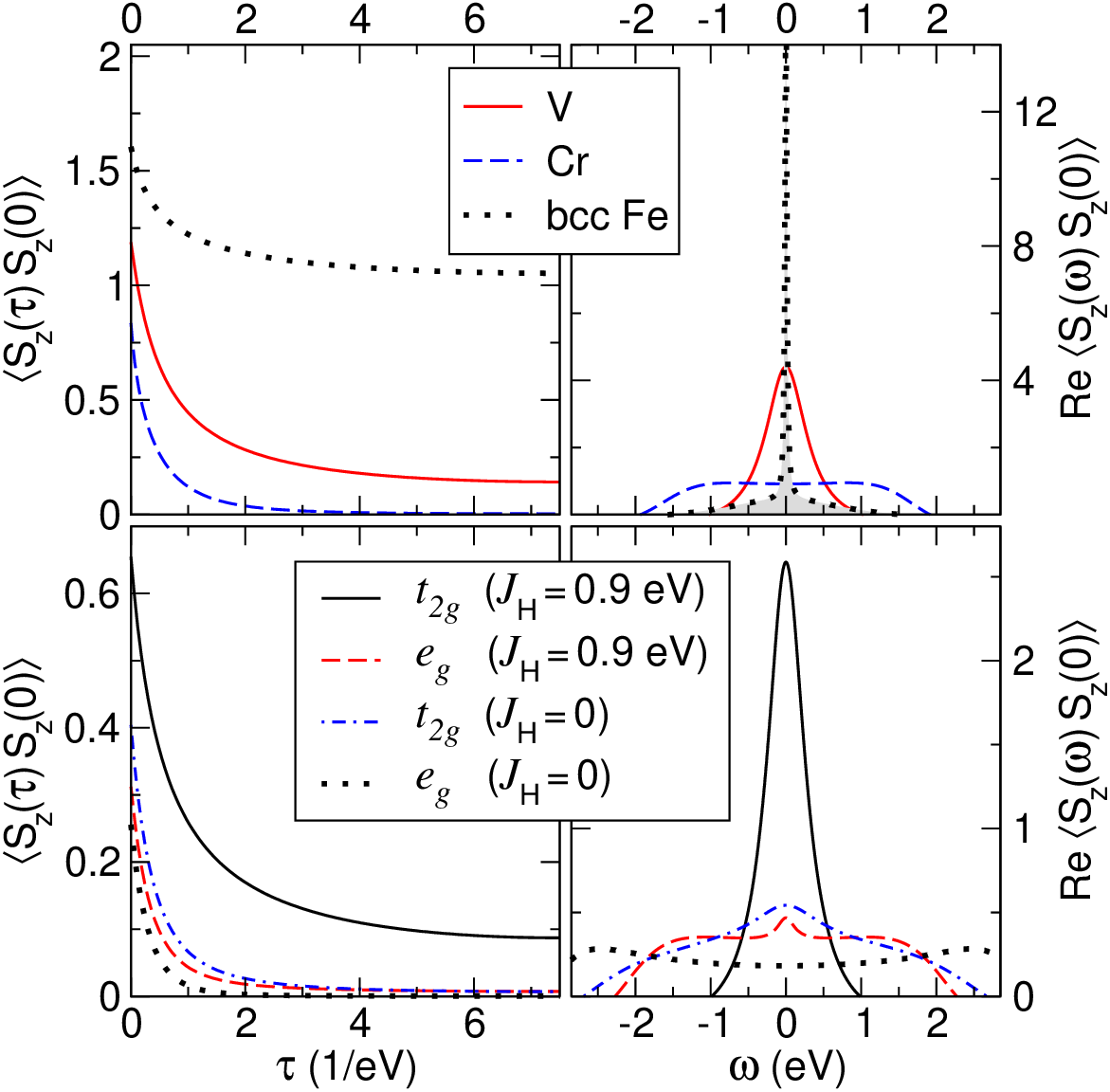}
\caption{
\label{Fig:correlators}
Local spin-spin correlation functions in the imaginary time $\tau$ (left panels) and real frequency $\omega$ (right panels) domains calculated by DFT+DMFT method for vanadium at temperature ${T=770}$~K 
with ${U=4}$~eV and ${J_{\rm H}=0.9}$~eV
in comparison with those of paramagnetic chromium~\cite{ourChromium} and bcc iron~\cite{OurAlphaBelozerovKatanin} (top panels).
Bottom panels: orbital-resolved spin-spin correlation functions for vanadium obtained within DFT+DMFT at ${T=770}$~K with ${U=4}$~eV and different values of $J_{\rm H}$.}
\end{figure}

\begin{figure}[t]
\centering
\includegraphics[clip=true,width=0.58\textwidth]{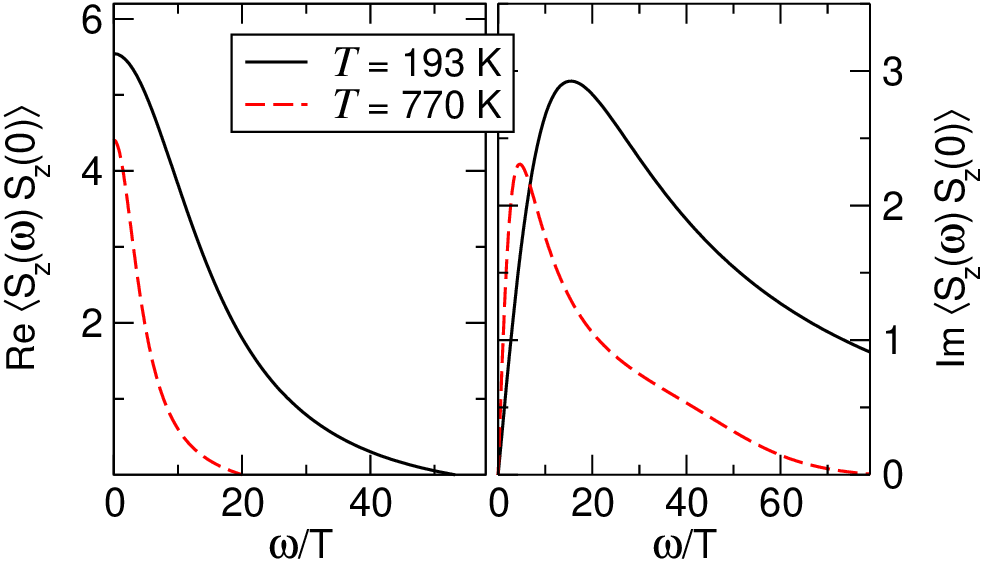}
\caption{
\label{Fig:correlators_two_T}
Real (left panel) and imaginary (right panel) parts of local spin-spin correlation functions for vanadium calculated by DFT+DMFT method
with ${U=4}$~eV and $J_{\rm H}=0.9$~eV
at two different temperatures.
}
\end{figure}

To clarify the degree of magnetic moment localization at various temperatures, we consider the local spin-spin correlation function ${\chi_{\rm spin}(\tau) = \langle S_z(\tau) S_z(0) \rangle}$ and its real-frequency counterpart
%
%
$\chi_{\rm spin}(\omega)$, obtained by Fourier transform of ${\chi_{\rm spin}(\tau)}$ to imaginary bosonic frequency followed by an analytical continuation to real frequency~$\omega$ using Pad\'e approximants~\cite{Pade}.
In the top panel of Fig.~\ref{Fig:correlators}, we present the obtained ${\chi_{\rm spin}(\tau)}$ and the real part of ${\chi_{\rm spin}(\omega)}$ at ${U=4}$~eV and ${J_{\rm H}=0.9}$~eV in comparison with those of paramagnetic chromium~\cite{ourChromium}, a canonical itinerant antiferromagnet, and bcc iron~\cite{OurAlphaBelozerovKatanin}, a metal with well-defined local magnetic moments~\cite{footnote1}.
%
%
One can see that ${\chi_{\rm spin}(\tau)}$ for vanadium has
an instantaneous average ${\langle S_z^2 \rangle = 1.2}$ corresponding to instantaneous spin $S$ close to~$3/2$.
With an increase of $\tau$, ${\chi_{\rm spin}(\tau)}$ for V and Cr decay faster than that for Fe. 
However, in contract to ${\chi_{\rm spin}(\tau)}$ for Cr, 
the one for V saturates to a finite value at ${\tau\to\beta/2}$, implying some localization of magnetic moments.

A more quantitative estimate of the degree of magnetic moment localization can be obtained from the dynamic local susceptibility ${\chi_{\rm spin}(\omega)}$, shown in the right panels of Fig.~\ref{Fig:correlators} and in Fig. \ref{Fig:correlators_two_T}.
In particular, the half-width of the peak of ${{\rm Re}\, \chi_\textrm{spin}(\omega)}$ at its half-height, which approximately coincides with the position of the peak of ${{\rm Im}\, \chi_\textrm{spin}(\omega)}$, is approximately equal to the inverse lifetime of local magnetic moments $h/\tau$ ~\cite{OurGamma,Toschi3}.
At $T=770$~K we obtain that lifetime of local magnetic moments in vanadium at ${U=4}$~eV and ${J_{\rm H}=0.9}$~eV
is about 20 times less than in iron, but about 5 times larger than in chromium. While at the temperatures $T>T_{\rm onset}$ we find $h/\tau\gtrsim T$, which corresponds to partial formation of local magnetic moments, at $T<T_{\rm onset}$ we have $h/\tau\gg T$, which reflects short lifetime of local magnetic moments due to their screening. The obtained lifetime of local magnetic moments in vanadium is also comparable to that determined from the position of the maximum of the imaginary part of local susceptibility in pnictides~\cite{Toschi}.
In addition, considering a lower ${U=2.3}$~eV leads to a decrease of local magnetic moment lifetime in vanadium by only 20\%.

The orbital-resolved spin-spin correlation functions for vanadium, shown in the bottom panel of Fig.~\ref{Fig:correlators}, indicate that the partially formed local magnetic moments originate only from states of $t_{2g}$ symmetry. This is in line with the larger density of $t_{2g}$ states at the Fermi level and obtained non-analytic behavior of the self-energy for $t_{2g}$ states. We also performed calculations with ${J_{\rm H} = 0}$, which resulted in almost complete disappearance of local magnetic moments (see Fig.~\ref{Fig:correlators}), confirming that their formation is governed by the Hund's rule coupling.

\begin{figure}[t]
\centering
\includegraphics[clip=true,width=0.56\textwidth]{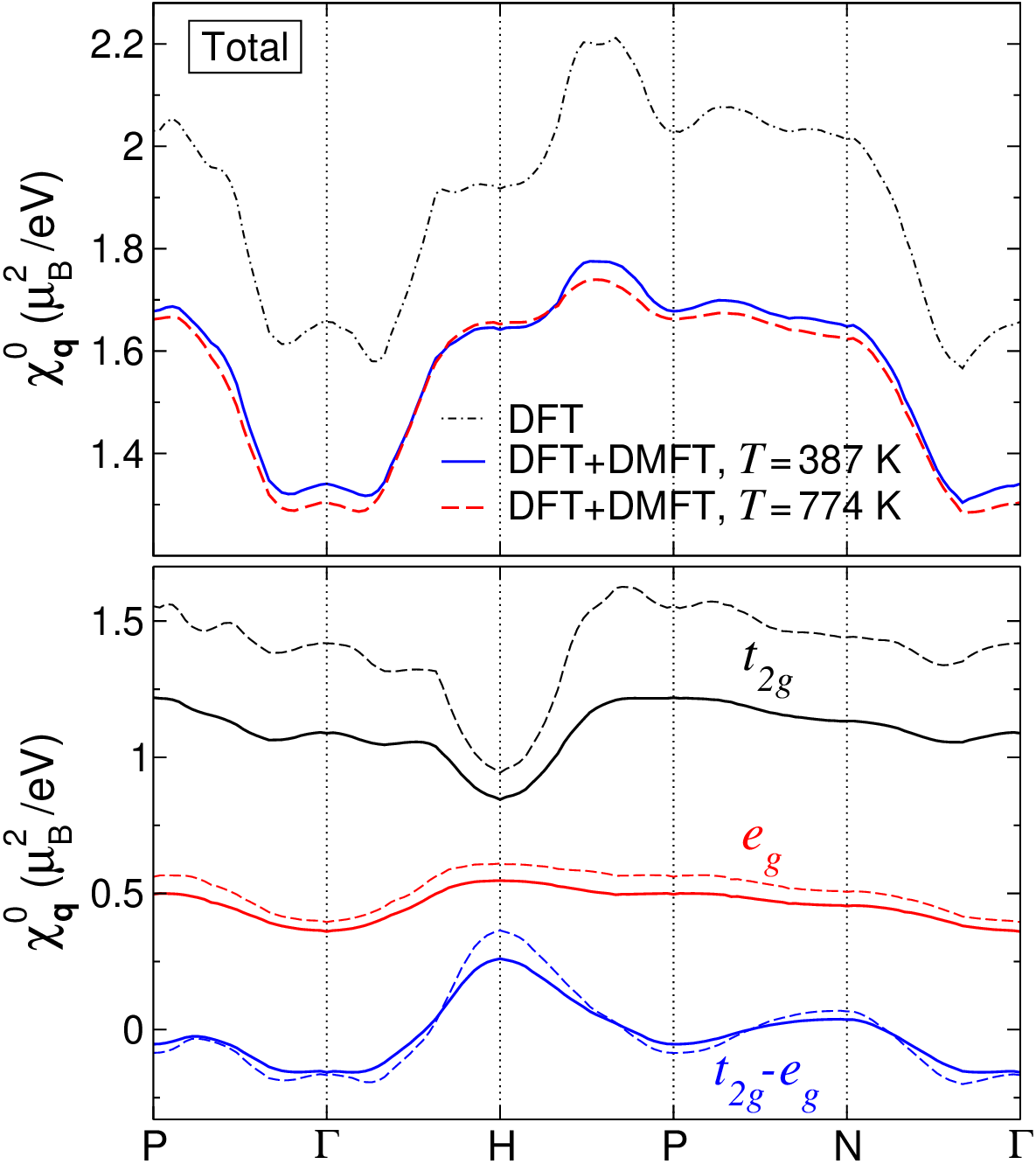}
\caption{
\label{fig:susc_irr}
Momentum dependence of the static magnetic susceptibility for $d$ states (top panel) and its orbital-resolved contributions (bottom panel) obtained within DFT and DFT+DMFT at ${U=4}$~eV and ${J_{\rm H}=0.9}$~eV. In the bottom panel, the DFT results are shown by dashed lines, while the DFT+DMFT ones at temperature ${T=774}$~K are drawn by solid lines.}
\end{figure}

To determine a preferable wave-vector of magnetic response, we compute the momentum dependence of the static magnetic susceptibility by considering the lowest-order contribution (i.e., particle-hole bubble) 
which does not account for the vertex corrections:
\begin{equation}
\chi_{\bf q}^{0} = -\frac{2\mu_B^2}{\beta} \sum_{{\bf k},\nu_n} \textrm{Tr}[G_{\bf k} (i\nu_n) G_{\bf k+q}(i\nu_n)],
\end{equation}
where 
$G_{\bf k}(i\nu_n)$ denotes the one-particle Green's function obtained using the Wannier-projected Hamiltonian matrix at momentum ${\bf k}$, $\nu_n$ are the fermionic Matsubara frequencies, $\beta$ is the inverse temperature and $\mu_B$ is the Bohr magneton.

In the top panel of Fig.~\ref{fig:susc_irr},
we present $\chi_{\bf q}^{0}$ for $d$ states obtained using non-interacting (DFT) and interacting (DFT+DMFT) Green's functions.
In DFT, $\chi_{\bf q}^{0}$ is found to have two closely located maxima at incommensurate wave vectors lying between H and P points of the Brillouin zone.
These maxima correspond to competing magnetic correlations. 
The overall shape of $\chi_{\bf q}^{0}$ is kept in DFT+DMFT,
but the maxima appear to be smeared
{by temperature and correlation effects.}
%
The maximum of the considered particle-hole bubble shows possible importance of incommensurate magnetic correlations, similarly to $\gamma$-iron \cite{OurGamma}, $\epsilon$-iron \cite{epsilon_fe}, and chromium \cite{ourChromium}.
%
As seen in the bottom panel of Fig.~\ref{fig:susc_irr}, the maxima of $\chi_{\bf q}^{0}$ occur due to ${t_{2g}}$ 
and partly due to mixed $t_{2g}$-$e_g$ 
contribution. 
%

\subsection{Rigid-band shift of the Fermi level  \label{Sect:model}}

\begin{figure}[b]
\centering
\includegraphics[clip=true,width=0.61\textwidth]{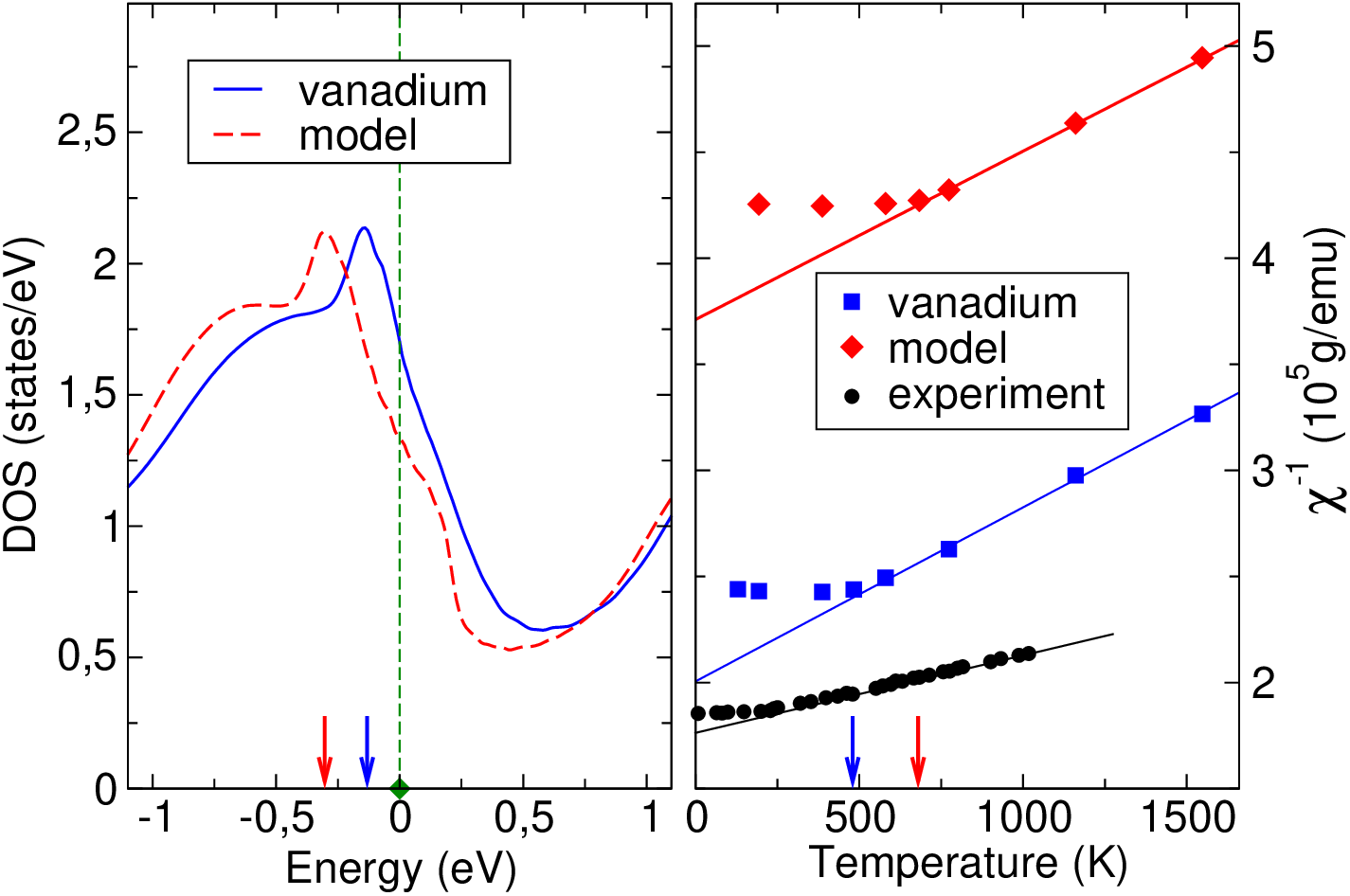}
\caption{
\label{Fig:model}
Total density of states (left panel) and inverse of uniform magnetic susceptibility (right panel) calculated by DFT+DMFT method for vanadium and a model obtained by rigid-band shift of the Fermi level with ${U=4}$~eV and ${J_{\rm H}=0.9}$~eV.
The DOS was computed at temperature of 387~K.
The experimental data are from Ref.~\cite{susc2_suzuki65}. 
The Fermi level in the left panel is at zero energy.
The straight lines in the right panel depict the least-squares fit to the linear dependence at high temperatures.}
\end{figure}

To establish whether the transition to Curie-Weiss behavior is related to the peak of DOS near the Fermi level, we perform a rigid-band shift of the Fermi level in DFT+DMFT calculations as shown in the left panel of Fig.~\ref{Fig:model}.
For this model, the value of the shift was chosen to approximately double the distance from the peak to the Fermi level. 
In particular, we increased the particle number by 0.25 in DMFT part,
that allowed us to shift the peak location from $-0.14$ to $-0.29$~eV with respect to the Fermi level at ${U=4}$~eV and ${J_{\rm H}=0.9}$~eV.

Inverse of the obtained uniform magnetic susceptibility for the considered model is shown in the right panel of Fig.~\ref{Fig:model} in comparison with 
that of vanadium.
One can see that the temperature of the onset of Curie-Weiss behavior $T_{\rm onset}\sim T_{\rm K}$ is increased by only 40--50\% due to the shift of the Fermi level.
{
This indicates that $T_{\rm onset}$ is controlled not only by the distance of the Fermi level to the peak, but also by other energy scales.
}
The obtained increase of $T_{\rm onset}$  may be also associated with a weakening of electronic correlation strength due to a lower DOS at the Fermi level.

%
%

\section{Conclusion} \label{sec:conclusions}

We have investigated the electronic and magnetic properties of vanadium by DFT+DMFT approach. 
The local dynamical correlations are found to be essential for a description of spectral and magnetic properties. 
In particular, the obtained DOS with Hubbard ${U=2.3}$~eV
is shown to agree well with experimental spectra.
The calculated uniform magnetic susceptibility in bcc structure qualitatively reproduces the transition from Pauli to Curie-Weiss behavior with heating at the temperature $T_{\rm onset}\sim 400-700$~K.
%
%
The latter is found to originate from 
the local spin correlations within $t_{2g}$ states associated with Hund's rule coupling.
The lifetime of local magnetic moments in vanadium is about 5~times larger than in paramagnetic chromium, a canonical
itinerant antiferromagnet, but is about 20 times less than in bcc iron, where the local moments are well formed. 
Therefore, Heisenberg-type models are not appropriate for a description of vanadium magnetism, which is mostly
of an itinerant origin. 

Our results demonstrate that the Curie-Weiss law and corresponding formation of local magnetic moments
in vanadium can be obtained in bcc structure upon heating. This shows that the
structural transition ~\cite{Bollinger2011} is not necessary to explain these phenomena, and possibly does not occur. The obtained behavior originates solely from the electronic degrees of freedom, and also does not require lattice (phonon) contribution.
%
%

The obtained self-energies have a quasiparticle shape, but, however, show a deviation from Fermi-liquid behavior above temperature ${T^*\sim 200}$~K, which is below
the calculated temperature of 
the onset of Curie-Weiss behavior 500--700~K. In particular, we obtain the non-analytic form of the self-energy of $t_{2g}$ states and 
almost linear temperature dependence of the quasiparticle damping at $T>T^*$, which may have important effect on the transport properties.
The quasiparticle mass enhancement factor ${m^*/m}$ of $e_g$ states is found to be in the range 1.4--1.9 depending on value of $U$. 
%
%
These results
characterize vanadium as a moderately correlated metal which is on the verge of the spin freezing transition.

Assuming Kondo temperature $T_K\sim T_{\rm onset}$, we find that the
%
local magnetic moments are likely almost screened at realistic temperatures, which provides their
short lifetime. 
With elevating temperature, the Curie-Weiss behavior of the uniform susceptibility is obtained, 
which is likely caused by the non-Fermi liquid properties of the electronic states and
respective formation of local magnetic moments. Calculation of the non-uniform susceptibility shows the presence of pronounced incommensurate magnetic correlations.

By considering a rigid-band shift of the Fermi level, we
have also demonstrated that the temperature of the onset of Curie-Weiss behavior 
depends on the position of peak in DOS, but not controlled solely by the position of the peak.
Analyzing the momentum dependence of magnetic susceptibility, we find its maximum 
at 
incommensurate wave-vectors, which result in incommensurate
magnetic correlations.
%
%

The agreement with experimental data may be further improved by (i) considering rotationally invariant form of on-site Coulomb interaction, which was shown to be important for a quantitative description of magnetic properties~\cite{Belozerov2013,Sangiovanni,Antipov2012} and (ii) taking into account non-local correlation effects beyond DMFT~\cite{beoyndDMFT}; the latter is exact in the limit of infinite coordination number.
However, at the moment such calculations are too computationally expensive especially at a temperature of several hundreds of kelvins. Therefore, these points should be addressed in further theoretical studies.

\begin{acknowledgments}
The DFT+DMFT calculations were supported by the
Russian Science Foundation (project 19-12-00012).
The calculations of the particle-hole bubble
were supported by the Ministry of Science and Higher Education of the Russian Federation (theme ``Electron" No. AAAA-A18-118020190098-5).
\end{acknowledgments}



\begin{thebibliography}{99}
\bibitem{footnote1}
The DFT+DMFT calculations of local spin-spin correlation functions for chromium and bcc iron were performed at the same temperature ${T=770}$~K as for vanadium. For bcc Fe 
the paramagnetic state was enforced by assuming spin-independent self-energy. Other computational details can be found in Refs.~\cite{ourChromium,OurAlphaBelozerovKatanin}.
\end{thebibliography}
\end{document}